\documentclass[aps,pra,twocolumn,groupedaddress,floatfix]{revtex4}

\usepackage{graphicx}
\usepackage{textcomp}
\usepackage{amsmath}

\begin{document}

\title{~\vspace{1.7cm}\\ Extremely asymmetrical scattering of electromagnetic
waves\\ in gradually varying periodic arrays}


\author{D. K. Gramotnev}

\author{T. A. Nieminen}

\affiliation{Centre for Medical and Health Physics,
School of Physical Sciences,
Queensland University of Technology, GPO Box 2434,
Brisbane Qld 4001, Australia}


\date{23rd December 1998}

\begin{abstract}
\vspace{-5.4cm}
\noindent
\hspace{-1.5cm}\textbf{Preprint of:}

\noindent
\hspace{-1.5cm}D. K. Gramotnev and T. A. Nieminen,

\noindent
\hspace{-1.5cm}``Extremely asymmetrical scattering of electromagnetic
waves in gradually varying periodic arrays'',

\noindent
\hspace{-1.5cm}\textit{Journal of Optics A: Pure and Applied Optics}
\textbf{1}, 635--645 (1999)

\hrulefill

\vspace{3.4cm}

This paper analyses theoretically and numerically the effect of varying
grating amplitude on the extremely asymmetrical scattering (EAS) of bulk
and guided optical modes in non-uniform strip-like periodic Bragg arrays
with stepwise and gradual variations in the grating amplitude across the
array. A recently developed new approach based on allowance for the
diffractional divergence of the scattered wave is used for this analysis.
It is demonstrated that gradual variations in magnitude of the grating
amplitude may change the pattern of EAS noticeably but not radically.
On the other hand, phase variations in the grating may result in a
radically new type of Bragg scattering---double-resonant EAS (DEAS).
In this case, a combination of two strong simultaneous resonances (one
with respect to frequency, and another with respect to the phase variation)
is predicted to take place in non-uniform arrays with a step-like phase and
gradual magnitude variations of the grating amplitude. The tolerances of EAS
and DEAS to small gradual variations in the grating amplitude are determined.
The main features of these types of scattering in non-uniform arrays are
explained by the diffractional divergence of the scattered wave inside and
outside the array.
\end{abstract}


\maketitle

\section{Introduction}

Extremely asymmetrical scattering (EAS) of waves in periodic arrays is a
new type of Bragg scattering that is realized when the scattered wave
propagates parallel or almost parallel to the boundary(ies) of a strip-like
periodic array~\cite{ref1,ref2,ref3,ref4,ref5,ref6,ref7,ref8,ref9,%
ref10,ref11}. This type of scattering is radically different from the
conventional Bragg scattering in periodic arrays. For example, EAS is
characterized by a strong resonant increase of the scattered wave
amplitudes inside and outside the array; the smaller the grating
amplitude, the larger the amplitudes of the scattered waves~\cite{ref2,ref6,%
ref7,ref8,ref9,ref10,ref11}. In addition, the incident and scattered waves
inside the array each split into three
waves~\cite{ref6,ref7,ref8,ref9,ref10,ref11}. Two of these scattered waves
and two of the incident waves inside the array are evanescent waves which
are localized near the array boundaries~\cite{ref6,ref7,ref8}. The third
scattered wave is a plane wave propagating at a grazing angle into the
array~\cite{ref6,ref7,ref8}.

Our recent publications~\cite{ref6,ref7,ref8} have demonstrated that
the physical reason for EAS is related to the diffractional divergence
of the scattered wave inside and outside the array. A new powerful
approach for simple analytical analysis of EAS, based on allowance for
this diffractional divergence, was introduced and
justified~\cite{ref6,ref7,ref8,ref9,ref10,ref11}. In
the case of bulk electromagnetic waves, this approach was shown to give
the same coupled mode equations as the dynamic theory of scattering that
was previously used for the theoretical
analysis of EAS~\cite{ref1,ref2,ref3,ref4,ref5,ref8}. However, the dynamic
theory of scattering is not suitable for analysis of EAS of guided and
surface waves on account of extremely awkward calculations involved, while
the new approach is readily applicable for all types of waves, including
surface and guided optical and acoustic modes~\cite{ref6,ref7,ref9,ref10}.

EAS has enormous potential for new important practical applications in the
development of novel optical communication devices (e.g. narrow-band optical
filters, resonators, couplers, switches, lasers, etc), optical sensors and
measurement techniques. For example, the strong resonant increase of the
scattered wave amplitude during EAS may result in high-quality EAS-based
resonators and high sensitivity of sensors and measurement techniques. New
non-collinear geometry of EAS can lead to the development of highly tuneable
optical and ultrasonic devices, and may also result in an improved side-lobe
structure of filtered signals. The possibility of concentration of the
scattered wave energy in narrowchannels can be used for amplification and
lateral compression of waves, as well as for effective coupling of a planar
waveguide and a fibre.

However, manufacturing EAS-based structures and devices will inevitably
be related with various imperfections of periodic arrays. Because of the
resonant character of EAS, it is possible to expect that at least some of
these imperfections may be crucial for the experimental observation and
practical use of this type of scattering. Thus, a very important
practical problem is to investigate theoretically the effect of array
imperfections on EAS, and determine the tolerance of the scattering. In
addition, non-uniform arrays are often manufactured on purpose. For example,
non-uniform chirped gratings can effectively compress pulses broadened due
to dispersion of optical modes in a slab or optical fibre~\cite{ref12}
(i.e. the dispersion can be compensated). Launching of solitons into
non-uniform nonlinear Bragg gratings can be significantly easier than
for uniform nonlinear Bragg gratings~\cite{ref13,ref14}. The side-lobe
structure of filtered signals can be noticeably improved (suppressed) by
using non-uniform gratings with slowly varying grating amplitude~\cite{ref15}.

There are also two other special reasons for using nonuniform arrays in the
case of EAS. Firstly, edge effects at the array boundaries may result in
noticeable undesirable energy losses in the scattered wave, caused by an
additional re-scattering of the scattered wave. These losses may be
noticeable because of the large amplitude of the scattered wave
propagating along the boundaries. Non-uniform arrays with gradually
increasing grating amplitude can be used for a substantial reduction
of this effect. Secondly, nonuniform arrays with varying grating amplitude
may result in a radically new type of EAS~\cite{ref10}, as well as cause
concentration of the wave energy within narrow channels inside the
array~\cite{ref11}. In addition, non-uniform arrays may result in a
significant reduction of the relaxation time to steady-state EAS.

Therefore, EAS of bulk or guided opticalwaves has been analysed for
non-uniform arrays with step-like variations of the grating
amplitude~\cite{ref10,ref11}. It was demonstrated that a step-like
variation in magnitude of the grating amplitude may result in a
noticeable re-arrangement of the intensity distribution of the scattered
wave inside the array~\cite{ref11}. The sensitivity of EAS to small uniform
and non-uniform (step-like) variations in the grating amplitude was also
determined~\cite{ref11}. Step-like phase variations in the grating inside
the array were shown to have a much stronger effect on EAS than the magnitude
variations~\cite{ref10}. In thin arrays, they resulted in a radically new
type of EAS double-resonant extremely asymmetrical scattering (DEAS) that
is characterized by a unique combination of two sharp simultaneous
resonances with respect to frequency and the phase variation in the
grating~\cite{ref10}.

However, non-uniform arrays with step-like variations of the grating
amplitude will hardly reduce edge effects at the array boundaries.
Moreover, the presence of additional interfaces at which the step-like
variations take place must increase undesirable edge effects and the
resultant energy losses. This is especially the case for DEAS, where
the scattered wave amplitude is exceptionally large~\cite{ref10}. Step-like
variations in the grating amplitude will also hardly improve the side-lobe
structure of processed (filtered) signals. Finally, manufacturing periodic
arrays usually results in imperfections characterized by slow (but not
step-like) variations in the grating amplitude (e.g. due to non-uniform
etching, photolithography, etc). However, EAS and DEAS in non-uniform
arrays with slow variations of the grating amplitude have not been
analysed previously, though such analysis is crucial for successful
development of new practical applications of these types of scattering.

Therefore, the aim of this paper is to analyse theoretically and numerically
the steady-state EAS and DEAS of bulk and guided electromagnetic waves in
non-uniform periodic arrays with slowly varying grating amplitude. Coupled
wave equations describing EAS and DEAS in arrays with slowly varying grating
amplitude are derived. The field structure in the scattered waves is
determined inside and outside the array by means of numerical solution
of the coupled wave equations. The tolerance of EAS and DEAS to small
slow variations in the grating amplitude (array imperfections) is also
investigated theoretically. The results are compared with those obtained
for EAS and DEAS in uniform and non-uniform arrays with step-like variations
of the grating amplitude.

\section{Coupled wave equations}

The structure under investigation and the geometry of scattering are
presented in figure 1. An incident wave is scattered in a strip-like
periodic array of width $L$. The period of the grating is assumed to be
constant in the whole array, while the grating amplitude can vary in phase
and/or in magnitude across the array, i.e. in the $x$-direction (figure 1).
We will consider only stepwise and/or slow (i.e. small at distances of the
order of the grating period) variations in the grating amplitude. In this
case, the grating amplitude can be assumed to be locally constant at each
point of the array, except for the values of the x-coordinate at which the
steplike variations take place (at $x = L_1$ in figure 1). The array is
assumed to be uniform along the $y$-axis, i.e. in the directions parallel
to the array boundaries.

\begin{figure}[hb]
\includegraphics[width=0.7\columnwidth]{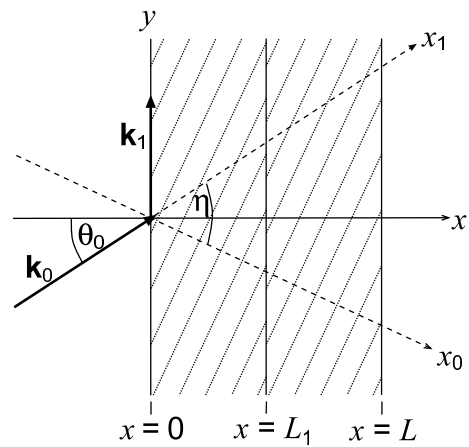}
\caption{Scheme for EAS in non-uniform strip-like periodic Bragg arrays
with varying grating amplitude and the total array width equal to $L$.}
\end{figure}

In sections 2--5 we analyse EAS and DEAS of bulk TE optical waves in
periodic Bragg arrays represented by a periodic variation of the dielectric
permittivity:
\begin{multline}
\epsilon_s =
\epsilon + \epsilon_1(x)\exp(\mathrm{i}\mathbf{q}\cdot\mathbf{r})
+ \epsilon_1^\ast(x)\exp(-\mathrm{i}\mathbf{q}\cdot\mathbf{r})\\
\shoveright{\text{if $0<x<L$},}\\
\shoveleft{\epsilon_s =\epsilon} \;\;\;\;\;\;\;\;\;
\text{if $x < 0$, or $x > L$},
\end{multline}
where the mean dielectric permittivity $\epsilon$ is the same in all parts
of the structure (inside and outside the array), the amplitude of the
grating $\epsilon_1(x)$ is small:
\begin{equation}
|\epsilon_1(x)|/\epsilon \ll 1,
\end{equation}
and $\mathbf{q}$ is the reciprocal lattice vector
($|\mathbf{q}| = 2\pi/\Lambda$, where $\Lambda$ is the grating period).
There is no dissipation of electromagnetic waves inside or outside the
array, i.e. $\epsilon$ is real and positive. The complex grating amplitude
$\epsilon_1(x)$ varies in magnitude and/or phase within the array.

A plane TE electromagnetic wave (with the electric field parallel to the
$z$-axis) is incident on the array at an angle $\theta_0$ (measured
counterclockwise from the $x$-axis---figure 1). We assume that the Bragg
condition is satisfied precisely:
\begin{equation}
\mathbf{k}_1 - \mathbf{k}_0 = - \mathbf{q}
\end{equation}
where $\mathbf{k}_0$ is the wavevector of the incident wave, $\mathbf{k}_1$
is parallel to the array boundaries (figure 1),
$|\mathbf{k}_1| = |\mathbf{k}_0| = k_0 = \omega\epsilon^{1/2}/c$, $\omega$
is the angular frequency, and $c$ is the speed of light in vacuum.

If condition (2) is satisfied for all values of $x$, then the amplitudes
of the incident and scattered waves vary slowly inside the array, i.e.
their variations at distances of about one wavelength are small as compared
with the values of these amplitudes. In this case the approximation of slowly
varying amplitudes is valid and only two waves---incident and scattered---need
to be taken into account inside and outside the array:
\begin{multline}
E(x) = E_0(x)\exp\{\mathrm{i}k_{0x}x + \mathrm{i}k_{0y}y -
\mathrm{i}\omega t\} + \\
E_1(x)\exp\{ \mathrm{i}k_0 y - \mathrm{i}\omega t\},
\end{multline}
where $E_0(x)$ and $E_1(x)$ are the slowly varying amplitudes of the
electric field in the incident and scattered waves, respectively,
$k_{0x} = k_0 \cos\theta_0$, and $k_{0y} = k_0 \sin\theta_0$.

It was shown previously~\cite{ref6,ref7,ref8,ref9,ref10,ref11} that in the
geometry of EAS, there are two opposing mechanisms determining the behaviour
of the scattered wave amplitude. On the one hand, the scattered wave
amplitude must increase along the direction of its propagation (i.e.
along the $y$-axis) due to scattering of the incident wave inside the
array. On the other hand, the scatteredwave amplitude must decrease along
the $y$-axis due to a significant diffractional divergence of this
wave~\cite{ref6,ref7,ref8,ref9,ref10,ref11}. The steady-state EAS occurs
when the increment in the scattered wave amplitude caused by the scattering
is exactly compensated by the decrement caused by the diffractional
divergence~\cite{ref6,ref7,ref8,ref9,ref10,ref11}. The new approach for the
theoretical analysis of EAS is based on the separate determination of each
of the two contributions to the scattered wave amplitude. The coupled wave
equations describing the steady-state EAS are then derived from the
comparison of these contributions~\cite{ref6,ref7,ref8,ref9,ref10,ref11}.

Here, we apply this approach to the considered case of EAS and DEAS in
non-uniform periodic arrays with slowly varying grating amplitude. During
the first step, we determine the contribution due to scattering,
disregarding the diffractional divergence. In this case the amplitude of
the scattered wave must increase along the $y$-axis. If the amplitudes of
the incident and scattered waves vary slowly inside the array, then at any
point in the array, the incident wave amplitude and the grating amplitude
can be considered to be locally constant. Thus the scatteringinduced
increments in the amplitude of the scattered wave along the direction of
its propagation at any point of the array are determined only by the local
values of the grating amplitude and the amplitude of the incident wave,
regardless of the type of Bragg scattering. Therefore, we can find these
increments by means of the conventional dynamic theory of scattering in
uniform arrays.

The coupledwave equations in the conventional dynamic theory of scattering
are well known~\cite{ref8,ref12,ref13,ref14,ref15}:
\begin{eqnarray}
\mathrm{d}E_0/\mathrm{d}x_0 & = & \mathrm{i}\Gamma_1 E_1,\\
\mathrm{d}E_1/\mathrm{d}x_0 & = & \mathrm{i}\Gamma_0 E_0,
\end{eqnarray}
where $\Gamma_0$ and $\Gamma_1$ are the coupling coefficients, and the
$x_0$-axis is parallel to the reciprocal vector of the grating (see figure 1).
Equation (6) gives the scattering-induced rate of changing the scattered wave
amplitude along the $x_0$-axis. The rate of changing amplitude of the
scattered wave along the direction of its propagation can then be obtained
from equation (6) by means of the simple substitution of
$\mathrm{d}x_0 = -\cos(pi/2-\eta+\theta_0)\mathrm{d}y
= -\sin(\eta-\theta_0)\mathrm{d}y$:
\begin{equation}
\left( \frac{\partial E_1}{\partial y} \right)_{\mathrm{scattering}} =
- \mathrm{i} \Gamma_0 E_0(x) \sin(\eta-\theta_0).
\end{equation}
where $\eta$ is the angle (measured counterclockwise) between the positive
$x_0$-direction and the wavevector of the incident wave (figure 1).

During the second step, we disregard the scattering and consider only the
diffractional divergence of the scattered wave (beam). In this case, the
rate of decreasing amplitude of the scattered wave can be derived by
substituting the entire scattered field
$E_\mathrm{sc} = E_1\exp(\mathrm{i}k_1 y - \mathrm{i}\omega t)$ into the
Helmholtz equation $\nabla^2 E_\mathrm{sc} + k_1^2 E_\mathrm{sc} = 0$:
\begin{equation}
\frac{\partial^2}{\partial x^2}E_1 + \frac{\partial^2}{\partial y^2}E_1
+ 2\mathrm{i}k\frac{\partial}{\partial y}E_1 = 0.
\end{equation}
Here, we can neglect the second-order derivative of the slowly varying
amplitude $E_1$ with respect to y as compared with the first-order
derivative. This gives us the parabolic equation of diffraction    
\begin{equation}
\left( \frac{\partial E_1}{\partial y} \right)_{\mathrm{divergence}} =
\frac{\mathrm{i}}{2k_1} \frac{\partial^2 E_1}{\partial x^2}
\end{equation}
that determines the rate of decreasing scattered wave amplitude along
the direction of its propagation due to the diffractional divergence.

As has already been mentioned, in the steady-state EAS, the contributions
to the scattered wave amplitude, caused by the scattering and diffractional
divergence, must exactly compensate each other. Therefore, the sum of rates
(7) and (9) must give zero. This condition results in the following equation:
\begin{equation}
\frac{\mathrm{d}^2 E_1(x)}{\mathrm{d}x^2} + K_0 E_0(x) = 0,
\end{equation}
where
\begin{equation}
K_0 = -2k_1 \Gamma_0 \sin(\eta-\theta_0).
\end{equation}

Note that in equations (8)--(11) we have deliberately used $k_1$ instead
$k_0$, even though $k_1 = k_0$ for bulk electromagnetic waves. This was done
to make the presented derivation valid for all types of waves, including
surface and guided modes, for which $k_1$ may be not equal to $k_0$
(e.g. for scattering of TE modes guided by a slab into TM modes of the
same slab).

Equation (10) is one of the coupled wave equations describing EAS (or DEAS)
in a non-uniform periodic array. The second equation is obtained from
equation (5). Recall that in the considered geometry of scattering (see
figure 1), the incident wave amplitude must depend only on the $x$-coordinate.
Therefore, substituting $\mathrm{d}x_0 = \mathrm{d}x_1 \cos\eta =
\mathrm{d}x \cos\eta / \cos\theta_0$ into equation (5) gives:
\begin{equation}
\frac{\mathrm{d} E_0(x)}{\mathrm{d}x} = \mathrm{i} K_1 E_1(x),
\end{equation}
where
\begin{equation}
K_1 = \Gamma_1 \cos\eta / \cos\theta_0.
\end{equation}

Equations (10) and (12) are the complete set of coupled wave equations for
EAS (or DEAS) of electromagnetic waves in a non-uniform periodic array with
varying grating amplitude. As has already been mentioned above, these coupled
wave equations are valid for the description of EAS and DEAS of all types of
waves, including surface and guided optical modes. The difference between,
for example, bulk and guided optical waves is only in different values of
the coupling coefficients $\Gamma_0$ and $\Gamma_1$. For bulk electromagnetic
waves~\cite{ref8}
\begin{equation}
\Gamma_0 = -\Gamma_1^\ast = -\epsilon_1^\ast\omega^2/[2c^2k_0\cos\eta]
\end{equation}
while for guided modes, these coefficients are determined using one of the
conventional dynamic theories of scattering for optical slab
modes~\cite{ref12,ref13,ref14,ref15} (see section 6 for more detail).

There is no coupling between incident and scattered waves outside the
array, i.e. at $x < 0$ and $x > L$. Therefore, the coupling coefficients
$K_{0,1}$ are equal to zero outside the array, and the coupled wave
equations (10) and (12) take the form:
\begin{equation}
\mathrm{d}^2E_1(x)/\mathrm{d}x^2 = 0; \;\;\;\;\;
\mathrm{d}E_0(x)/\mathrm{d}x = 0.
\end{equation}

One can easily see that the coupled wave equations (10), (12) and (15) in
the case of non-uniform arrays with slowly varying grating amplitude appear
to have exactly the same form as for uniform arrays~\cite{ref6,ref7,ref8,ref9}.
However, for non-uniform arrays, the coupling coefficients $K_1$ and $K_0$
in equations (10) and (12) are dependent on the $x$-coordinate because the
grating amplitude $\epsilon_1$ depends on $x$. This results in substantial
difficulties with analytical solution of these equations. Therefore, in
sections 3 and 4, we present results of numerical analysis of the coupled
wave equations (10) and (12) for periodic arrays with several typical
dependences of the grating amplitude on the $x$-coordinate. The numerical
method and boundary conditions used are described in the appendix.

\section{EAS in non-uniform arrays}

In this section, we analyse numerically the effect of slowly varying
grating amplitude on EAS of bulk TE electromagnetic waves in periodic
Bragg arrays. The results obtained for the non-uniform arrays are compared
with those for EAS in uniform arrays~\cite{ref8}. We also investigate
numerically the tolerance of EAS to small gradual variations (imperfections)
in the grating amplitude.

\begin{figure}[hb]
\includegraphics[width=\columnwidth]{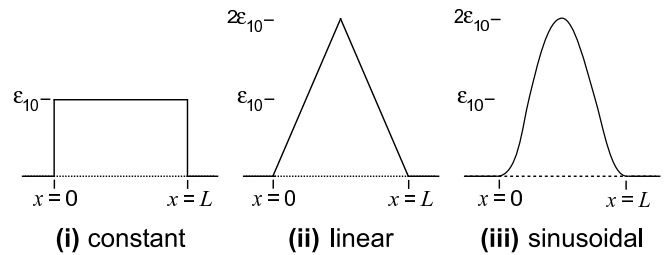}
\caption{Three different profiles of the grating amplitude inside a
periodic array: (i) constant grating amplitude (uniform array),
(ii) linearly varying grating amplitude (non-uniform array), and
(iii) sinusoidally varying grating amplitude (non-uniform array).
The mean grating amplitudes are the same.}
\end{figure}

\begin{figure*}[ht]
\includegraphics[width=0.9\textwidth]{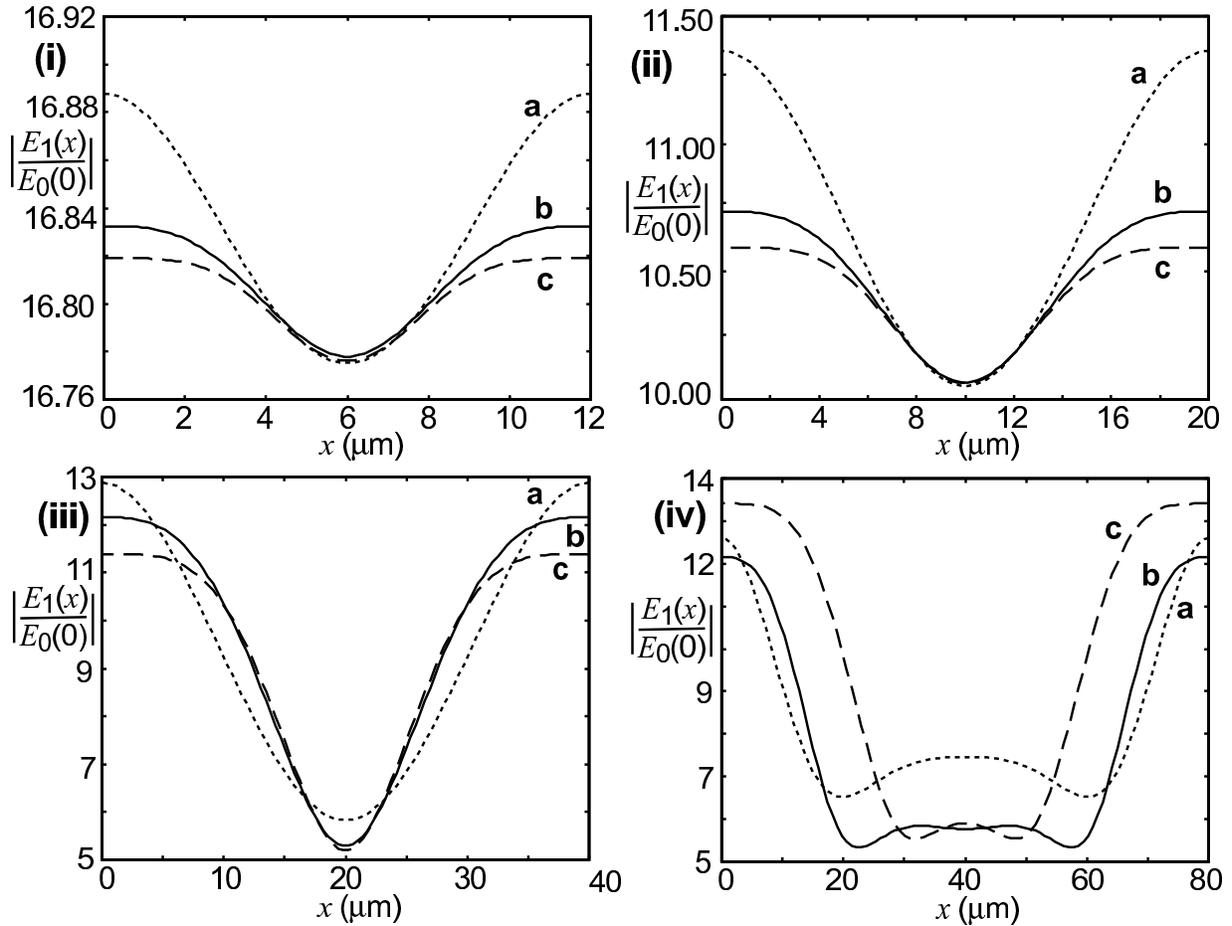}
\caption{The dependences of the relative scattered wave amplitudes on
distance from the front array boundary inside the arrays with
$\epsilon_{10}= 5\times 10^{-3}$, $\epsilon = 5$, $\theta_0 = \pi/4$, the
wavelength in vacuum $\lambda = 1$\,{\textmu}m (the grating has a period
of $0.58$\,{\textmu}m and is inclined at the angle of $\pi/8$ to the front
array boundary); (i) $L = 12$\,{\textmu}m, (ii) $L = 20$\,{\textmu}m,
(iii) $L = 40$\,{\textmu}m, (iv) $L = 80$\,{\textmu}m. Curves (a): uniform
array with constant grating amplitude $\epsilon_{10}$---figure 2(i).
Curves (b): non-uniform array with the linear dependence of
$\epsilon_1(x)$---figure 2(ii). Curves (c): non-uniform array with the
sinusoidal dependence of $\epsilon_1(x)$---figure 2(iii).}
\end{figure*}

For example, consider three different types of dependences of the grating
amplitude on the $x$-coordinate inside the array---figures 2(i)--(iii).
In figure 2(i), the grating amplitude is constant throughout the array,
i.e. we consider a uniform array of width $L$ and the grating amplitude
$\epsilon_{10}$. EAS in such arrays was investigated previously for
bulk~\cite{ref8} and guided~\cite{ref6,ref9} optical waves. Mathematically,
the profile of the grating amplitude in such an array is given by the
equations:
\begin{equation}
\epsilon_1(x) =
\begin{cases}
\epsilon_{10} & \text{if $0<x<L$},\\
0 & \text{otherwise}.
\end{cases}
\end{equation}

The second of the considered arrays is a non-uniform array with a linear
dependence of magnitude of the grating amplitude on the $x$-coordinate in
such a way that the grating amplitude is zero at the front and rear
boundaries of the array (figure 2(ii)):
\begin{equation}
\epsilon_1(x) =
\begin{cases}
4 \epsilon_{10} x/L & \text{if $0<x<L/2$},\\
4 \epsilon_{10} (L-x)/L & \text{if $L/2<x<L$},\\
0 & \text{otherwise}.
\end{cases}
\end{equation}
The gradient of changing grating amplitude in (17) is chosen such that
the mean grating amplitude in the non-uniform array is the same as in
the uniform array described by equations (16). Thus the integral
\begin{equation}
\int_0^L \epsilon_1(x)\mathrm{d}x
\end{equation}
must be the same for both the dependences presented by equations (16) and
(17). This condition is necessary, because EAS is strongly dependent on the
grating amplitude~\cite{ref6,ref7,ref8,ref9}. If, for example,
$\epsilon_1(x)$ increased gradually from zero to $\epsilon_{10}$, the mean
grating amplitude would be noticeably smaller than that of the uniform array,
resulting in a significant increase in the scattered wave amplitude. If we
assume that the mean grating amplitudes are the same for the uniform and
nonuniform arrays of the same widths (i.e. integral (18) is the same for
both the arrays), then we will be able to investigate only the effects of
non-uniformity of the grating on EAS.

The third of the considered arrays is characterized by sinusoidal
variations of the grating amplitude so that the grating amplitude is
again zero at the front and rear array boundaries (figure 2(iii)):
\begin{equation}
\epsilon_1(x) =
\begin{cases}
\epsilon_{10}[ 1 - \cos(2\pi x/L) ] & \text{if $0<x<L$},\\
0 & \text{otherwise}.
\end{cases}
\end{equation}
Here, the mean grating amplitude is again the same as for arrays (16) and
(17), i.e. integral (18) is the same for all three dependences (16), (17)
and (19). The relationship between the maxima of the grating amplitudes in
arrays (16), (17) and (19) can be written as
\begin{multline}
\max\{ \epsilon_1(x) \}_\mathrm{linear} =
\max\{ \epsilon_1(x) \}_\mathrm{sinusoidal}\\ =
2 \max\{ \epsilon_1(x) \}_\mathrm{stepwise} \equiv 2\epsilon_{10}.
\end{multline}

The selection of non-uniform arrays in the form of (17) and (19) is not
arbitrary. Analysis of arrays with zero grating amplitude at the front
and rear boundaries (as for arrays in figures 2(ii) and 2(iii)) is
practically important because such arrays may significantly improve
the side-lobe structure of filtered signals~\cite{ref15} and substantially
reduce unwanted edge effects at the array boundaries.

Figure 3 presents the dependences of the scattered wave amplitudes on the
$x$-coordinate inside periodic arrays (16), (17) and (19). It can be seen
that for all array widths, slowly varying grating amplitude does not
introduce radical changes in the pattern of scattering. It can also be
seen that the effect of slowly varying grating amplitude on amplitudes
of the scattered waves is noticeably smaller for narrow arrays. For
example, for arrays of 12\,{\textmu}m thickness (figure 3(i)),
the difference between the scattered wave amplitude $E_1$ in the
non-uniform array with sinusoidal variation of the grating amplitude (19)
and the scattered wave amplitude in the uniform array (16) is less than 0.4\%.
For linear dependence (17) this difference is even smaller (figure 3(i)).
At the same time, it can reach 10--15\% in arrays of 40\,{\textmu}m width
(figure 3(iii)), and over 25\% in arrays of 80\,{\textmu}m width
(figure 3(iv)).

This can be explained by the diffractional divergence of the scattered wave.
In narrow arrays, the diffractional divergence effectively smooths out
variations in the amplitude of the scattered wave, caused by different
values of the grating amplitude. For such arrays, the scattered wave
amplitudes are determined (to a very good accuracy) by an average value
of the grating amplitude (figures 3(i) and (ii)). However, the diffractional
divergence is effective only within a limited distance. This is because when
the scattered wave spreads along the x-axis (due to the diffractional
divergence) from one part of the array to another, it experiences
rescattering in the grating. As a result, the wave can diverge along the
$x$-axis only until it is re-scattered by the grating, i.e. it can spread
only within a finite distance (see also~\cite{ref10}). This distance must
increase with decreasing mean grating amplitude, because the efficiency of
re-scattering decreases with decreasing grating amplitude. It should also
increase with increasing gradient $\mathrm{d}\epsilon_1(x)/\mathrm{d}x$,
because in this case variations of the amplitude of the scattered wave along
the wavefront will increase, resulting in stronger diffractional divergence.
Typically, the distance through which the scattered wave can spread inside
the array before being re-scattered is about several
$\gamma^{-1}$~\cite{ref10}, where $\gamma = (K_0 K_1)^{1/3}$.

In our examples (see equations (16), (17), (19) and figures 2 and 3),
the critical array width is about $\sim 3\gamma_\mathrm{av}^{-1}$, where
$\gamma_\mathrm{av}$ is the average value of $\gamma$ in the considered
arrays. Therefore, if $L < 3\gamma_\mathrm{av}^{-1}$, then the scattered
wave amplitude is very well determined by the average value of the grating
amplitude inside the array (figures 3(i) and 3(ii)). If the array width
$L > 3\gamma_\mathrm{av}^{-1}$, then the scattered wave amplitudes inside
a non-uniform array are significantly affected by local values of the
grating amplitude, and the pattern of scattering differs from that for
the uniform array with the average grating amplitude (figures 3(iii) and
3(iv)).

So far, we have analysed non-uniform arrays with significant variations
of the grating amplitude. Another practically important problem is the
effect on EAS of small grating imperfections, e.g. variations in the
grating amplitude, which are inevitable during manufacture of periodic
arrays. Therefore, in this paper,we analyse tolerance of EAS to small
variations of the grating amplitude across the array (i.e. in the
$x$-direction).

\begin{figure}[ht]
\includegraphics[width=\columnwidth]{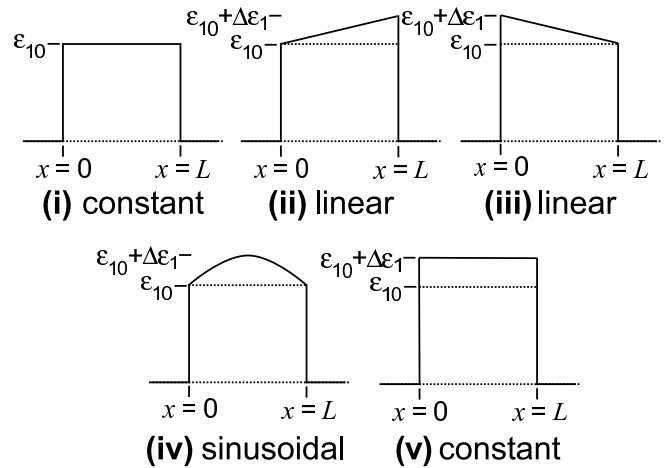}
\caption{Five different profiles of the grating amplitude with small
variations inside a periodic array: (i) constant grating amplitude
$\epsilon_{10}$ (uniform array), (ii) linearly increasing grating
amplitude (non-uniform array), (iii) linearly decreasing grating
amplitude (non-uniform array), (iv) sinusoidally varying grating
amplitude (non-uniform array), and (v) constant grating amplitude
$\epsilon_{10} + \Delta\epsilon_1$ (uniform array).}
\end{figure}

Figure 4 presents three different types of small and slow variations
of the grating amplitude on the $x$-coordinate in an initially uniform
periodic array. These are linearly increasing (figure 4(ii)), decreasing
(figure 4(iii)) and sinusoidal (figure 4(iv)) dependences of the grating
amplitude with the minimum and maximum values of the grating amplitude
equal to $\epsilon_{10}$ and $\epsilon_{10} + \Delta\epsilon_1$,
respectively, where $\Delta\epsilon_1 \ll \epsilon_{10}$.

\begin{figure*}[ht]
\includegraphics[width=0.9\textwidth]{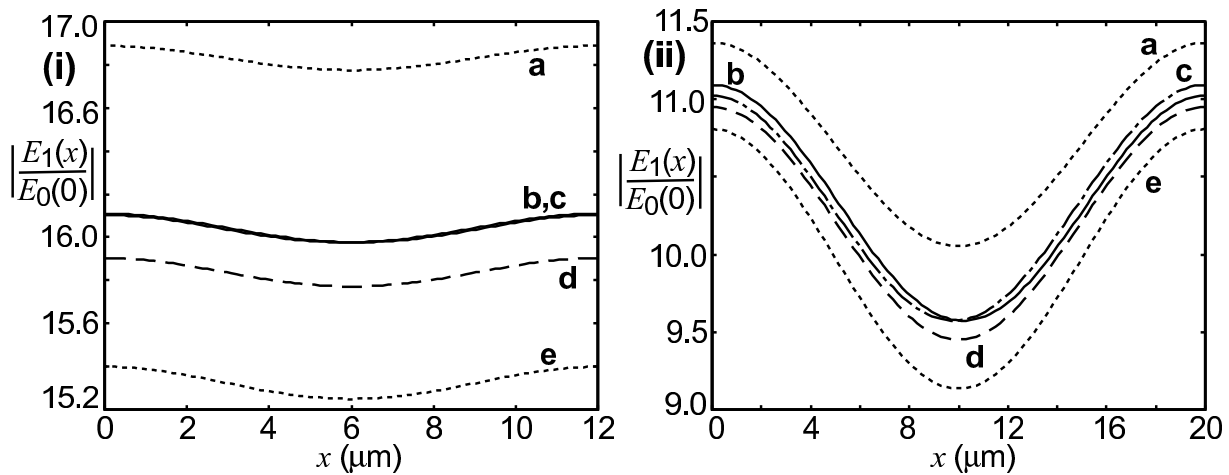}
\caption{The dependences of the relative scattered wave amplitudes on
distance from the front array boundary inside the arrays (figure 4) with
$\epsilon_{10} = 5\times 10^{-3}$, $\Delta\epsilon_1 = \epsilon_{10}/10$,
$\epsilon = 5$, $\theta_0 = pi/4$, the wavelength in vacuum
$\lambda = 1$\,{\textmu}m (the grating has a period of $0.58$\,{\textmu}m
and is inclined at an angle of $\pi/8$ to the front array boundary);
(i) $L = 12$\,{\textmu}m, (ii) $L = 20$\,{\textmu}m. Curves (a): uniform
arrays with constant grating amplitude $\epsilon_{10}$---figure 4(i).
Curves (b): non-uniform arrays with linearly increasing
$\epsilon_1(x)$---figure 4(ii). Curves (c): non-uniform arrays with
linearly decreasing $\epsilon_1(x)$---figure 4(iii). Curves (d):
non-uniform arrays with sinusoidal dependence of $\epsilon_1(x)$---figure
4(iv). Curves (e): uniform arrays with constant grating amplitude
$\epsilon_{10} + \Delta\epsilon_1$---figure 4(v).}
\end{figure*}

The results of the numerical analysis of the effect of small and slow
variations of the grating amplitude on the scattered wave amplitude
inside the array are presented in figures 5(i) and 5(ii) for
$\Delta\epsilon_1 = \epsilon_{10}/10 = 5\times10^{-4}$ for two
different array widths: (i) $L = 12$\,{\textmu}m, and (ii)
$L = 20$\,{\textmu}m. It can be seen that curves (b) and (c) in
figure 5(i) coincide with each other. This again reflects the fact that
for narrow arrays, the specific dependence of small gradual variations of
the grating amplitude does not affect the scattered wave amplitudes that
are dependent only on the average values of the grating amplitude in the
array. Curve (d) in figure 5(i) differs from curves (b) and (c). However,
this difference is related not with the sinusoidal profile of the dependence
of the grating amplitude on the $x$-coordinate (figure 4(iv)), but with the
slightly different value for the mean grating amplitude in this array. If
the mean grating amplitudes were the same for all three arrays shown in
figures 4(ii)--(iv), then curves (b)--(d) in figure 5(i) would have been
indistinguishable. Curves (a) and (e) correspond to the uniform arrays
with the
grating amplitudes $\epsilon_{10}$ (figure 4(i)) and
$\epsilon_{10} + \Delta\epsilon_1$ (figure 4(v)), respectively.

Figure 5(ii) shows that as the array width increases, the shape of the
dependence of the grating amplitude on the xcoordinate becomes significant.
For example, for the array width of 20\,{\textmu}m, the scattered wave
amplitudes are slightly different for arrays with increasing and
decreasing grating amplitude (figures 4(ii) and (iii))---see curves (b) and
(c) in figure 5(ii). If the array width is increased to 40 or more
\,{\textmu}m (i.e. $L > 3\gamma_\mathrm{av}^{-1}$), then the scattered
wave amplitude at any point inside the array is approximately the same
as it would have been in the uniform array of the same width and with the
grating amplitude taken from the point of observation in the non-uniform
array. That is, the scatteredwave amplitudes are determined by local values
of the grating amplitude. Note that this statement is correct only for slow
and small variations (imperfections) of the grating amplitude (figure 4).
It is not valid for arrays where the grating amplitude tends to zero
somewhere in the array or at its boundaries (see figures 2 and 3). If
the above statement was valid for such arrays, this would have meant that
the scattered wave amplitude in parts of the array with
$\epsilon_1 \rightarrow 0$ must increase to infinity, and due to the
diffractional divergence, it must then be infinite in all other parts of
the array, which is not possible.

Increase in $\Delta\epsilon_1$ results in variations of the scattered
wave amplitudes that are to a very good approximation directly proportional
to $\Delta\epsilon_1$. Therefore, the graphs in figure 5 can be used for
evaluation of the scattered wave amplitudes for any other value of
$\Delta\epsilon_1 \ll \epsilon_1$.

\section{DEAS in arrays with varying grating amplitude}

As has been seen from the analysis of the previous section, magnitude
variations of the grating amplitude may result in noticeable, but not
dramatic, changes in the field distribution inside the array. For narrow
arrays, such changes can even be neglected altogether. However, if we
consider phase variations in the grating, the situation may change very
radically, and especially for narrow arrays.

\begin{figure}[hb]
\includegraphics[width=\columnwidth]{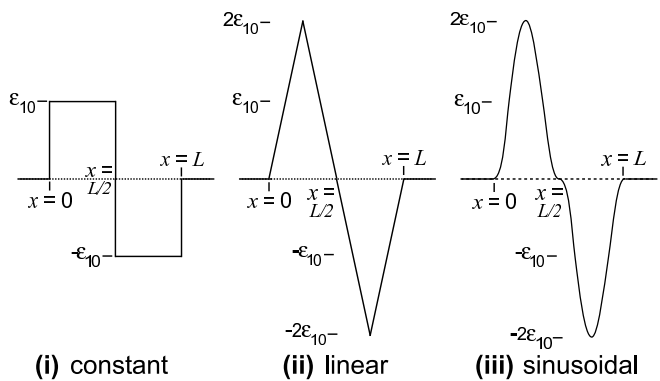}
\caption{Three different profiles of the grating amplitude with a stepwise
phase variation $\phi\approx 180^\circ$ at $x = L/2$: (i) constant
magnitude of the grating amplitude, (ii) linearly varying grating
amplitude, and (iii) sinusoidally varying grating amplitude.}
\end{figure}

For example, consider three different types of dependences of the grating
amplitude on the $x$-coordinate inside a non-uniform
array---figures 6(i)--(iii). In figure 6(i), the magnitude of the
grating amplitude is constant throughout the array, but the phase of
the grating experiences a stepwise
variation $\phi\approx 180^\circ$ in the middle of the array, i.e. the sign
of the grating amplitude $\epsilon_1(x)$ changes when crossing the interface
between the two sections of the array at $x = L_1 = L/2$---see figure 6(i)
and figure 1. In all other parts of the array the phase of the grating is
not changing (figure 6(i)). Thus we have a non-uniform array consisting of
two joint uniform arrays with different phases of the gratings. In the
general case with arbitrary stepwise phase shift $\phi$ between the gratings
in the joint arrays, the dependence of the grating amplitude on the
$x$-coordinate is given by the equations:
\begin{equation}
\epsilon_1(x) =
\begin{cases}
\epsilon_{10} & \text{if $0<x<L/2$},\\
\epsilon_{10}\exp(\mathrm{i}\phi) & \text{if $L/2<x<L$},\\
0 & \text{otherwise}.\\
\end{cases}
\end{equation}
If $\phi\approx 180^\circ$, then equations (21) give the dependence of
the grating amplitude presented in figure 6(i).

The second considered type of non-uniform arrays is characterized by a
linear dependence of magnitude of the grating amplitude on the $x$-coordinate
in such a way, that the amplitude of the grating is zero at the front and
rear array boundaries, as well as at the interface $x = L_1 = L/2$
(figure 6(ii)). There is also a stepwise variation in the phase of the
non-uniform grating at $x = L_1 = L/2$, which is equal to
$\phi\approx 180^\circ$. Thus we have two joint non-uniform arrays with
linearly varying magnitude of the grating amplitude and constant phase of
the grating in each of the joint arrays (figure 6(ii)). If the stepwise
variation of the phase between the two joint arrays is arbitrary, then
the dependence of the grating amplitude on the $x$-coordinate is given by
\begin{equation}
\epsilon_1(x) =
\begin{cases}
4\epsilon_{10}x/L & \text{if $0<x\le L/4$},\\
4\epsilon_{10}(L/2-x)/L & \text{if $L/4<x\le L/2$},\\
4\epsilon_{10}\exp(\mathrm{i}\phi)(x-L/2)/L & \text{if $L/2<x\le 3L/4$},\\
4\epsilon_{10}\exp(\mathrm{i}\phi)(L-x)/L & \text{if $3L/4<x<L$},\\
0 & \text{otherwise},\\
\end{cases}
\end{equation}
where the gradient of changing magnitude of the grating amplitude is
chosen so that the mean magnitude of the grating amplitude in the array
described by equations (22) is the same as for the array described by
equations (21). That is,
\begin{equation}
\int_0^{L/2} |\epsilon_1(x)|\mathrm{d}x =
\int_{L/2}^L |\epsilon_1(x)|\mathrm{d}x = G,
\end{equation}
where the constant $G$ is the same for both the dependences (21) and (22).
The reasons for using this condition are the same as the reasons for
integral (18) being the same for the arrays with the grating amplitude
given by equations (16), (17), and (19).

The third of the considered arrays with a phase variation in the grating
is presented in figure 6(iii). It is characterized by sinusoidal variations
of magnitude of the grating amplitude inside the joint arrays in such a way
that this magnitude is again zero at the boundaries of the periodic
array and at the interfaces $x = L_1 = L/2$. The phase of the grating is
again constant inside each of the joint arrays, but experiences a stepwise
variation $\phi\approx 180^\circ$ at $x = L_1 = L/2$ (where the magnitude
of the grating amplitude is zero)---figure 6(iii). If the phase variation
$\phi$ is arbitrary, then such non-uniform arrays can be described by the
equations:
\begin{equation}
\epsilon_1(x) =
\begin{cases}
\epsilon_{10}[1 - \cos(4\pi x/L)] & \text{if $0<x<L/2$},\\
\epsilon_{10}\mathrm{e}^{\mathrm{i}\phi}\{1- \cos[2\pi(2x/L)]\} &
\text{if $L/2<x<L$},\\
0 & \text{otherwise}.\\
\end{cases}
\end{equation}
where condition (23) is again satisfied.

\begin{figure}[ht]
\includegraphics[width=\columnwidth]{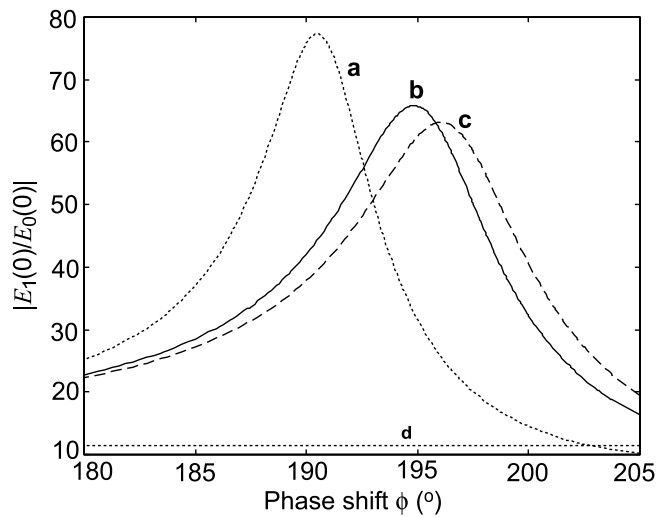}
\caption{The dependences of the relative scattered wave amplitudes at the
front boundary on the phase shift $\phi$ for the non-uniform arrays
described by equations (21), (22), (24) with $\epsilon_{10}=5\times 10^{-3}$,
$\epsilon = 5$, $\theta_0 = \pi/4$, $L= 20$\,{\textmu}m, and the wavelength
in vacuum $\lambda = 1$\,{\textmu}m. Curve (a): non-uniform array with
constant magnitude of the grating amplitude. Curve (b): non-uniform array
with the linear dependence of $\epsilon_1(x)$. Curve (c): non-uniform array
with the sinusoidal dependence of $\epsilon_1(x)$. Curve (d): uniform
array with $\phi = 0$.}
\end{figure}

\begin{figure*}[ht]
\includegraphics[width=0.9\textwidth]{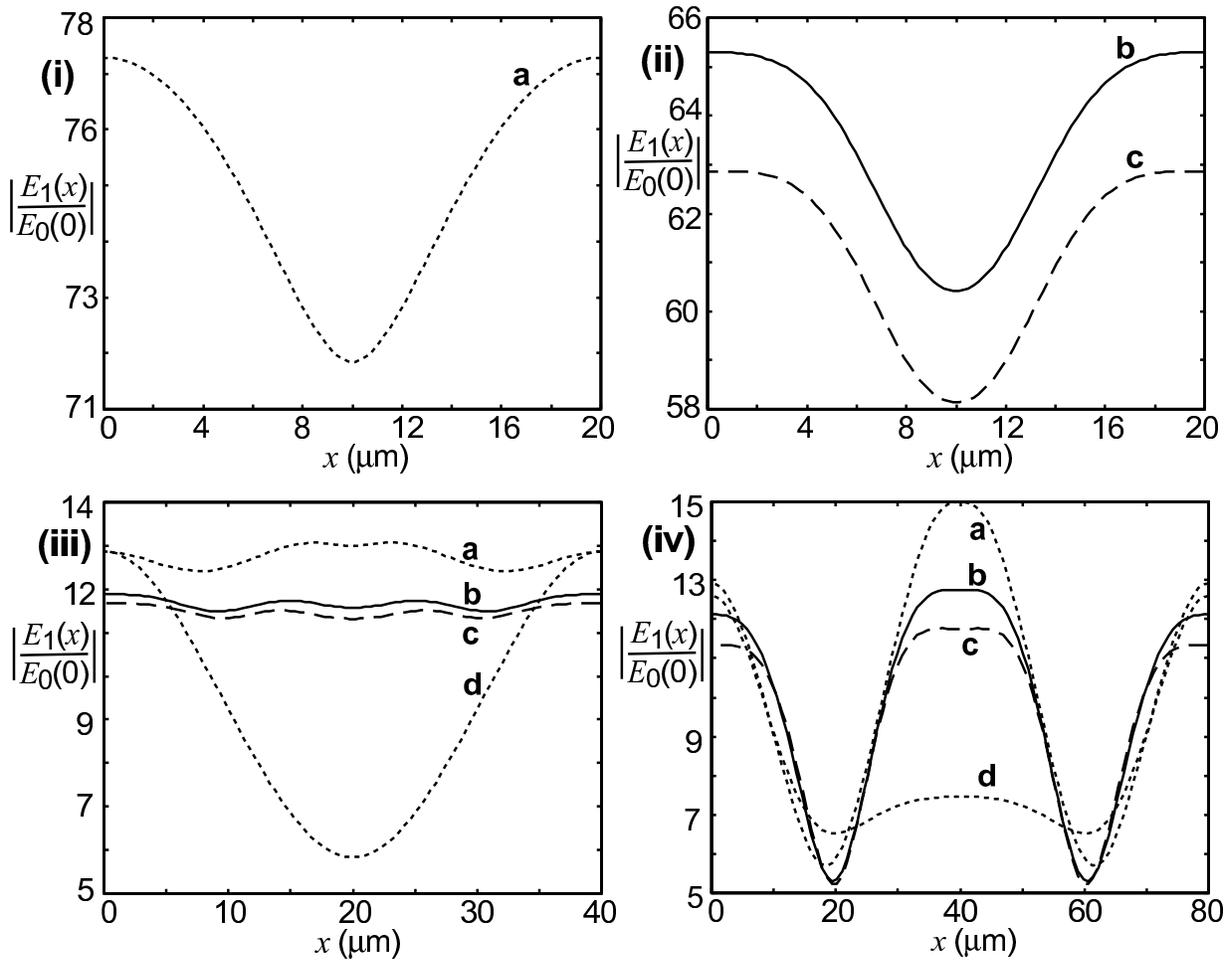}
\caption{The dependences of the relative scattered wave amplitudes on
distance from the front array boundary inside the non-uniform arrays
described by equations (21), (22), (24) with $\epsilon_{10}=5\times 10^{-3}$,
$\epsilon = 5$, $\theta_0 = \pi/4$, the wavelength in vacuum
$\lambda = 1$\,{\textmu}m; (i) and (ii) $L = 20$\,{\textmu}m, (iii)
$L = 40$\,{\textmu}m, (iv) $L = 80$\,{\textmu}m. Curves (a): non-uniform
arrays with constant magnitude of the grating amplitude (see equations
(21)); (i) $\phi = 190.4^\circ$ , (iii) $\phi = 178.6^\circ$ , (iv)
$\phi = 157.6^\circ$. Curves (b): non-uniform arrays with the linear
dependence of $\epsilon_1(x)$ (see equations (22)); (ii) $\phi = 194.7^\circ$,
(iii) $\phi = 182.5^\circ$, (iv) $\phi = 133.8^\circ$. Curves (c):
non-uniform arrays with the sinusoidal dependence of $\epsilon_1(x)$
(see equation (24)); (ii) $\phi = 195.9^\circ$, (iii) $\phi = 184.6^\circ$,
(iv) $\phi = 129.4^\circ$. Curves (d): uniform arrays with $\phi = 0$
and constant grating amplitude $\epsilon_1$.}
\end{figure*}

The dependences of the ratio of the amplitudes of the scattered and incident
waves at the front array boundary on the phase shift $\phi$ in non-uniform
arrays (21), (22) and (24) with $L = 20$\,{\textmu}m and
$\epsilon_{10} = 5\times 10^{-3}$ are presented in figure 7 for bulk TE
electromagnetic waves that are incident onto the arrays at the angle
$\theta_0 = \pi/4$. This figure demonstrates that when the value of the
phase shift is relatively close to $pi$, then the amplitude of the scattered
wave at the front array boundary very strongly (resonantly) increases as
compared with the conventional EAS (cf curves (a)--(c) with curve (d) in
figure 7). Thus we have two simultaneous resonances in the structure. One
of these resonances occurs at a resonant frequency (wavelength) that is
determined by the Bragg condition (3). This resonant is obviously common
for all types of Bragg scattering. In the case of EAS, it results in a
strong increase of the scattered wave amplitude see curve (d) in figure 7.

The second resonance takes place at an optimal (resonant) phase shift
in a non-uniform array. This resonance occurs on the background of an
already resonantly large scattered wave amplitude that is typical for
the conventional
EAS. As a result, the scattered wave amplitude increases many times
further as compared with the incident wave amplitude (see curves (a)--(c)
in figure 7). This effect was called DEAS~\cite{ref10}.

Curve (a) in figure 7 corresponds to the non-uniform array with the
grating amplitude given by equations (21). This is the case of DEAS that
was analysed previously in \cite{ref10} for bulk electromagnetic waves.
Curves (b) and (c) demonstrate that similar strong DEAS also occurs in
non-uniform arrays with varying magnitude of the grating amplitude, e.g.
in arrays with the grating amplitude given by equations (22) and (24).
It can be seen that the maxima of the scattered wave amplitude at the
front array boundary are about 15\% lower than for curve (a), and resonant
values of the phase variation are shifted to larger values, i.e. to about
$195^\circ$ as compared with $190.4^\circ$ for the array with constant
magnitude of the grating---curve (a). This demonstrates that slowly
increasing grating amplitude may result in noticeable changes of resonant
values of $\phi$ as compared with DEAS in arrays with constant magnitude
of the grating amplitude. On the other hand, curves (b) and (c) are very
similar, and this indicates that there is no significant difference in DEAS
in non-uniform arrays with linearly and sinusoidally varying grating
amplitude. Thus DEAS proves to be fairly insensitive to the particular
profile of the slowly varying grating amplitude inside the array.

Physically, the differences between curves (a)--(c) in figure 7 are
related to the fact that if the grating amplitude has the profile described
by equations (22) and (24) (see also figures 6(ii) and 6(iii)), then the
amplitude of the grating near the middle of the array (where the phase
variation takes place) is very small, and it is smaller for the sinusoidal
dependence (24). Thus, the efficiency of re-scattering in this area is low.
However, this is the region near the interface with the phase variation,
where the diffractional divergence of the scattered wave and its
re-scattering in the array result in increasing the incident wave
amplitude, which in turn gives rise to larger scattered wave
amplitudes~\cite{ref10}. Therefore, the amplitude of the scattered wave
must be smaller in arrays with the grating amplitude that decreases in
the middle of the array (i.e. near the interface with the phase variation
in the grating). Thus the resonance with respect to phase variation is
broader and weaker for curves (b) and (c), than for curve (a) in figure 7.
Similarly, the resonance for curve (c), corresponding to sinusoidal
variations of the grating amplitude, is slightly weaker and broader
than that for the linear dependence (curve (b)).

The dependences of the scattered wave amplitude on the $x$-coordinate
inside the non-uniform arrays (21), (22) and (24) of different widths
are presented in figure 8. Here, the values of the phase shift $\phi$ are
chosen so that to give maximal scattered wave amplitude at the interface
$x = L/2$ where the step-like phase variation occurs. It can be seen that
as the array width decreases from 40--20\,{\textmu}m, the scattered wave
amplitude increases dramatically (more than five times). This feature is
typical of DEAS in arrays with constant and varying magnitude of the
grating amplitude (see also \cite{ref10}). As can be seen from figure 8,
$L = L_c \approx 40$\,{\textmu}m is a critical array width, because
strong DEAS takes place only if $L < 40$\,{\textmu}m. This critical
array width is the same for all the arrays (21), (22) and (24), regardless
of whether the magnitude of the grating amplitude is varying or not. As has
already been mentioned, the critical array width is determined by the
distance within which the scattered wave may spread along the $x$-axis
before being re-scattered by the grating~\cite{ref10}. The average grating
amplitudes in arrays (22) and (24) are the same as in array (21) (see
condition (23)), and therefore this distance is the same for the analysed
arrays.

If $L > L_c$ (figure 8(iv)), then the amplitude of the scattered wave
at the front and rear boundaries of the periodic arrays are approximately
the same as in the case of EAS in the uniform array, because the
diffractional divergence of the scattered wave from the opposite side
of the array makes only negligible contribution to the scattering. However,
in the middle of the array near the interface with the phase variation, we
observe noticeable maxima of the scattered wave amplitudes for all three
considered arrays. The width
of these maxima is the same and equal (as expected) to the critical array
width. The maximum corresponding to array (24) is the smallest, while the
maximum corresponding to the array with constant magnitude of the grating
amplitude is the biggest figure 8(iv). This is again due to the fact that
the scattering near the interface $x = L_1$ in arrays (22) and (24) is
impeded by the small grating amplitude.

\begin{figure}[ht]
\includegraphics[width=\columnwidth]{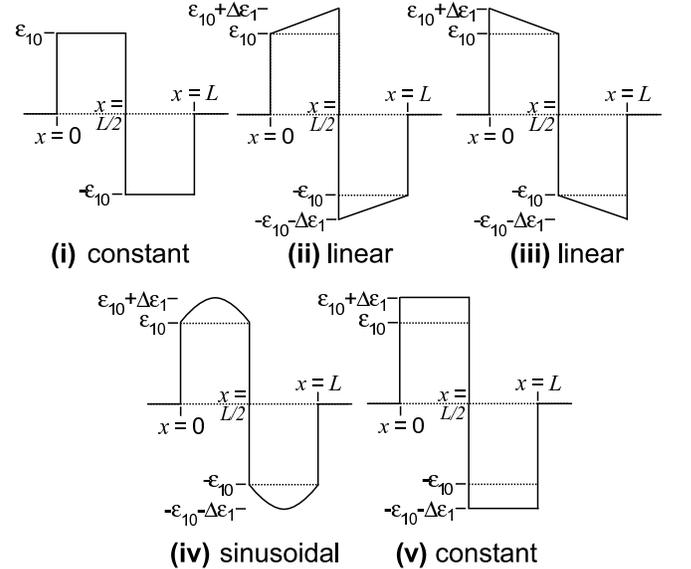}
\caption{Five different profiles of the grating amplitude inside
non-uniform periodic arrays. The grating amplitude has small gradual
magnitude variations, and a stepwise phase variation $\phi\approx 180^\circ$
at $x = L/2$.}
\end{figure}

\begin{figure}[ht]
\includegraphics[width=\columnwidth]{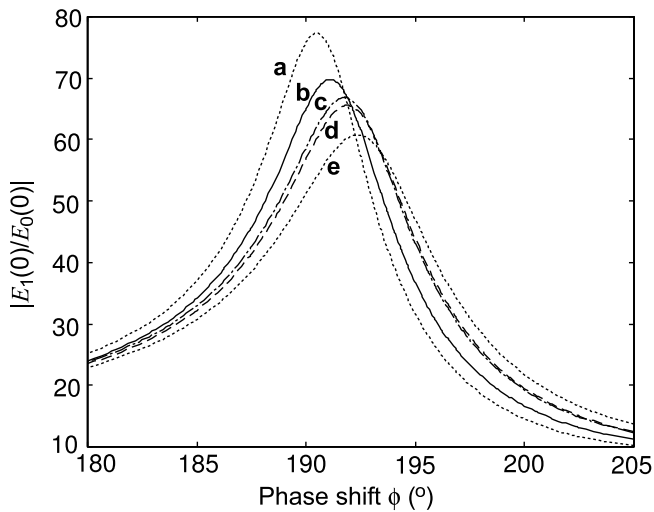}
\caption{The dependences of the relative scattered wave amplitudes at
the front array boundary on the phase shift $\phi$ for the non-uniform
arrays presented in figure 9 with $\epsilon_{10} = 5\times 10^{-3}$,
$\Delta\epsilon_1 = \epsilon_{10}/10$, $\epsilon = 5$, $\theta_0 = \pi/4$,
$L = 20$\,{\textmu}m, and the wavelength in vacuum $\lambda = 1$\,{\textmu}m.
Curves (a)--(e) correspond to the non-uniform arrays represented by figures
9(i)--(v), respectively.}
\end{figure}

The effect of small and gradual linear and sinusoidal variations of
magnitude of the grating amplitude (figure 9) on DEAS is presented in
figure 10 for $\Delta\epsilon_1 = \epsilon_{10}/10 = 5\times 10^{-4}$
and $L = 20$\,{\textmu}m. Note that unlike EAS in non-uniform arrays with
small variations of the grating amplitude (figure 5), DEAS appears to be
more sensitive to the particular shape of the dependence
of $\epsilon_1(x)$ on $x$---cf curves (b) and (c) in figure 10. The
sensitivity of DEAS to small imperfections of the array is also noticeably
stronger than for EAS (the difference between curves (e) and (a) in figure
10 reaches up to $\sim 50$\%). This is understandable because the resonance
during DEAS is noticeably sharper (due to the combination of the two
simultaneous resonances) than the resonance during EAS.

\section{Guided electromagnetic modes}

It was mentioned in section 2 that equations (5)--(13) and (15) are
valid for all types of waves, including guided optical modes. This is
because the approach based on allowance for the diffractional divergence
of the scattered wave, which led to the coupled wave equations (10) and
(12), is readily applicable for analysis of EAS of guided modes in a slab
with a periodically corrugated boundary. However, in this case the coupling
coefficients $\Gamma_0$ and $\Gamma_1$ are no longer given by equations
(14), but are determined in one of the following modern theories of mode
coupling in corrugated optical waveguides: boundary perturbation
theory~\cite{ref16}, mode matching theory~\cite{ref17},
local-normal-mode theory~\cite{ref18}, or direct approximate solution
of the wave equation~\cite{ref19}. These theories are correct for any type of
polarization (TE and TM) of slab modes. Therefore, the results of this
paper are also correct for any type of guided slab modes.

The geometry of EAS and DEAS of slab modes is again presented by figure 1.
In this case the plane of the figure is the plane of a slab, and the
periodic structure is represented by a periodic corrugation of a slab
boundary. If we consider a non-uniform periodic groove array with varying
grating amplitude, then the corrugation is given by the equations
\begin{equation}
\xi =
\begin{cases}
d + \xi_1(x)f(x_0) & \text{for $0<x<L$};\\
d & \text{for $x<0$ or $x>L$},
\end{cases}
\end{equation}
where $d$ is the thickness of the guiding slab, $\xi_1(x)$ is the varying
grating amplitude (corrugation amplitude), $f(x_0)$ is an arbitrary
periodic function with a period of $2\pi/q$, with $\max\{|f(x_0)|\} = 1$,
and a mean value of zero. Dissipation is neglected and all media in
contact are isotropic.

For the approximation of slowly varying amplitudes to be valid, the
corrugation must be small as compared with the grating period:
\begin{equation}
|\xi_1| q/(2\pi) \ll 1.
\end{equation}
This inequality is similar to condition (2) for bulk waves.

If the corrugation is non-sinusoidal (the function $f(x_0)$ is not a sine
or cosine), it can be expanded into the Fourier series
\begin{equation}
f(x_0) = \sum_{p=-\infty}^{+\infty} f_p \exp(\mathrm{i}pq_x x +
\mathrm{i}pq_y y) = \sum_{p=-\infty}^{+\infty} f_p\exp(\mathrm{i}pq x_0).
\end{equation}
In this case, only two complex conjugate terms (similar to equations
(1)) satisfying the Bragg condition
\begin{equation}
\mathbf{k}_1 - \mathbf{k}_0 = - p\mathbf{q}
\end{equation}
must be taken into account. Here, $\mathbf{q}$ is the reciprocal lattice
vector of the grating, which is parallel to the $x_0$-axis---see figure 1.

The vectors $\mathbf{k}_0$ and $\mathbf{k}_1$ are the wavevectors of the
incident and scattered guided modes. Note that in the case of guided modes,
$|\mathbf{k}_0 = |\mathbf{k}_1|$ only for TE$_n$--TE$_n$ or TM$_n$--TM$_n$
scattering ($n$ is the order of the mode). If EAS is related with
polarization change (e.g. TE$_n$--TM$_n$ scattering), or with change of
the mode order (e.g. TE$_n$--TE$_m$ scattering with $n \ne m$), then
$|\mathbf{k}_0 \ne |\mathbf{k}_1|$. That is why in equations (8)--(11) we
used $k_1$ even though for bulk waves $k_1$ is equal to $k_0$.

Therefore, all the speculations, derivations and results obtained in
sections 3--5 are applicable for EAS and DEAS of guided optical modes
in periodic groove arrays with varying corrugation amplitude if
$\epsilon_1(x)$, $\epsilon_{10}$, and $\Delta\epsilon_1$ are replaced by
$\xi_{1p}(x) = \xi_1(x)f_p$, $\xi_{10p} = \xi_{10}f_p$, and
$\Delta\xi_{1p} = \Delta\xi_1 f_p$, respectively. For example, consider
EAS or DEAS of an incident zeroth order TE slab mode into a scattered
zeroth order TE slab mode in the structure: vacuum---GaAs slab (with
permittivity 12.25)---AlGaAs substrate (with permittivity 10.24); a slab
thickness of $d = 0.6$\,{\textmu}m, angle of incidence $\theta_0 = \pi/4$,
the wavelength in vacuum $\lambda = 1.5$\,{\textmu}m, and the corrugation
is assumed to be sinusoidal, i.e. $f(x_0) = \sin(ax_0)$. If the corrugation
amplitude in this structure $\xi_1$ is given by the equation
$\xi_1 = 3.8\epsilon_1$ ({\textmu}m), where $\epsilon_1$ is the grating
amplitude used in sections 3--5, then the coupling coefficients $K_0$
and $K_1$ are the same as for the bulk TE waves in sections 3--5. Therefore,
all the graphs from these sections are valid for the scattering of the
guided slab modes in the above structure with the same (as for bulk waves)
array widths see sections 3--5.

\section{Conclusions}

In this paper, we have analysed theoretically and numerically the effect
of the grating amplitude, slowly varying across a non-uniform periodic
array, on EAS and DEAS of bulk and guided optical waves. The main
features of EAS in DEAS in such non-uniform arrays were explained by
the diffractional divergence of the scattered wave.

In particular, it was shown that the pattern of EAS in narrow arrays
with gradually varying magnitude of the grating amplitude is almost
exactly the same as for the uniform array with the same width and
grating amplitude equal to the average amplitude of the grating in
the nonuniform array. On the other hand, in thick arrays, the effect
of gradual variations in the grating amplitude on the scattered field
distribution was demonstrated to be rather noticeable. At the same time,
the main feature of EAS---the strong resonant increase in the scattered
wave amplitude---is typical for all considered non-uniform arrays. This
will allow use of non-uniform periodic arrays with gradually increasing
grating amplitude for suppression of edge effects that are expected to
be unusually strong during EAS, and for improving the side-lobe structure
of the scattered signals.

The tolerance of EAS to small gradual variations (imperfections) in
the grating amplitude inside the arrays was determined. It was shown
that EAS in narrow arrays is sensitive only to variations of the mean
value of the grating amplitude, while the particular shape of the
dependence of the grating amplitude on distance from the array boundaries
does not matter. In contrast, in wide arrays, the scattered wave amplitude
has been demonstrated to depend mainly on local values of the grating
amplitude.

It was also shown that DEAS which was previously analysed in arrays with
constant magnitude of the grating amplitude~\cite{ref10} also occurs in
non-uniform arrays with varying grating amplitude, though it appears to
be more sensitive to the particular profile of the grating amplitude than
EAS. Two strong simultaneous resonances in DEAS result in a much greater
increase in the scattered wave amplitude for the same amplitude of the
grating than during EAS. This is the reason why edge effects must be much
more significant for DEAS than for EAS, and the use of arrays with gradually
varying amplitude is even more important in this case. The sensitivity of
DEAS to small grating imperfections was shown to be noticeably stronger
than that of EAS, which is related to the much stronger resonance taking
place during DEAS.

The approach for the analysis used in this paper, is based on allowance
for the diffractional divergence of the scattered wave. Unlike the dynamic
theory of scattering~\cite{ref1,ref2,ref3,ref4,ref5,ref8} that is
applicable only for the analysis of EAS of bulk electromagnetic waves,
the approach used is directly applicable for the analysis of EAS and DEAS
of all types of waves, including bulk, guided and surface optical and
acoustic waves in uniform and non-uniform periodic Bragg arrays. Therefore,
the obtained results are also applicable to EAS and DEAS of Rayleigh surface
acoustic waves in periodic groove arrays. In this case we only need to use
in equations (11) and (13) the appropriate coupling coefficients $\Gamma_0$
and $\Gamma_1$ from the dynamic theory of scattering of Rayleigh waves in
periodic groove arrays~\cite{ref15}.

The obtained results will be important for development of new EAS- and
DEAS-based structures and devices in optical and acoustic signal-processing,
communication, instrumentation and sensor design.

\section*{Acknowledgment}

The authors gratefully acknowledge financial support for this work
from the Australian Research Council.

\appendix*

\section{}

The numerical solution of equations (10) and (12) was carried out using
the following boundary conditions at the array boundaries:
\begin{eqnarray}
E_0|_{x=0} & = & E_{00}, \nonumber\\
(\mathrm{d}E_1/\mathrm{d}x)_{x=0} & = & 0,\\
(\mathrm{d}E_1/\mathrm{d}x)_{x=L} & = & 0, \nonumber
\end{eqnarray}
where $E_{00}$ is the amplitude of the incident wave at $x < 0$. The
scattered wave outside the array is represented (similarly to uniform
arrays~\cite{ref6,ref7,ref8,ref9} by two plane waves propagating parallel
to the array boundaries one on each side of the array (these waves are the
solutions to equations (15)). Their amplitudes, $A_1$ for $x < 0$, and $A_2$
for $x > L$, are determined from two other conditions
\begin{equation}
A_1 = E_1|_{x = 0+0}, \;\;\;\; A_2 = E_1|_{x=L-0}
\end{equation}
that are actually independent of equations (A.1).

The condition $|E_0|_{x=L-0} = |E_{00}|$, representing the energy
conservation in the steady-state EAS, will be satisfied automatically
and we do not need to take it into account (see
also~\cite{ref6,ref7,ref8,ref9}). Since we neglect edge effects at the array
boundaries, the conditions implying the continuity of the derivative
$\mathrm{d}E_0/\mathrm{d}x$ across these boundaries are not taken into
account (this is the usual approximation in modern theories of Bragg
scattering\cite{ref6,ref7,ref8,ref9,ref10,ref11,ref15,ref16,ref17,ref18,%
ref19}). The grating amplitude inside the array is assumed to vary either
slowly, or very quickly (stepwise variations). These variations are taken
into account in the function $\epsilon_1(x)$ in the coupling coefficients
$K_{0,1}$. Therefore, we do not need to take into account boundary
conditions at interfaces with step-wise variations in the grating amplitude
inside the array (figure 1).

For convenience, the coupled wave equations (10) and (12) are rewritten
as a set of three linear first-order differential equations:
\begin{eqnarray}
\mathrm{d}E_0/\mathrm{d}x - \mathrm{i}K_1(x)E_1(x) & = & 0,\nonumber \\
\mathrm{d}E_1/\mathrm{d}x - E_2(x) & = & 0,\\
\mathrm{d}E_2/\mathrm{d}x + K_0(x)E_0(x) & = & 0.\nonumber
\end{eqnarray}

This set of differential equations can be approximated by a set of
$3 \times (N-1)$ finite difference equations (FDEs) on a set of $N$ points
$\{x_1,x_2,...,x_N\}$:
\begin{multline}
E_0(x_n) - E_0(x_{n-1}) + \frac{x_n - x_{n-1}}{4}[ E_0(x_n) + E_0(x_{n-1}) ]\\
\times [ \mathrm{i}K_1(x_n)E_1(x_n) + \mathrm{i}K_1(x_{n-1})E_1(x_{n-1}) ]
= 0\\
\shoveleft{E_1(x_n) - E_1(x_{n-1}) + \frac{x_n - x_{n-1}}{4}[ E_1(x_n) +
E_1(x_{n-1}) ]}\\
\times [ E_2(x_n) + E_2(x_{n-1})] = 0,\\
\shoveleft{E_0(x_n) - E_0(x_{n-1}) + \frac{x_n - x_{n-1}}{4}[ E_0(x_n) +
E_0(x_{n-1}) ]}\\
\times [ - K_0(x_n)E_2(x_n) - K_0(x_{n-1})E_2(x_{n-1}) ] = 0
\end{multline}
where $n = 2,3,...,N$. Since the amplitudes $E_0$ and $E_1$ are constant
outside the array, the points $x_1$ and $x_N$ are chosen to
correspond to the front and rear boundaries of the array, i.e.
$x_1 = 0$ and $x_N = L$. Then boundary conditions (A.1) can be written as:
\begin{equation}
E_0(x_1) = E_{00}, \;\;\; E_2(x_1) = 0, \;\;\; E_2(x_N) = 0.
\end{equation}
Equations (A.4) and (A.5) form a set of $3N$ linear algebraic equations
with $3N$ variables $E_0(x_1), E_0(x_2),...,E_0(x_N)$, $E_1(x_1), E_1(x_2),
...,E_1(x_N)$, $E_2(x_1), E_2(x_2),...,E_2(x_N)$, which was solved numerically
for different profiles of varying grating amplitude inside the array (see
sections 3--5).



\end{document}